\documentclass[journal=jacsat,manuscript=article,layout=twocolumn]{achemso}
\setkeys{acs}{keywords = true}
\usepackage[T1]{fontenc} 
\usepackage{graphicx}% Include figure files
\graphicspath{ {./Figures/} }
\usepackage{float}
\usepackage{dcolumn}% Align table columns on decimal point
\usepackage{bm}% bold math
\usepackage{tablefootnote} %table footnote
\usepackage{threeparttable} %table footnote 2nd attempt WORKING
\usepackage{xcolor}
\usepackage{soul}
\usepackage[export]{adjustbox}
\usepackage{flushend}
\usepackage{multicol}
\usepackage{amsmath}
\newcommand{\Squeeze}{\!\!\!\!}
\newcommand{\squeeze}{\!\!}
\usepackage{soul}
\usepackage{hyperref}

\author{Ransell D'Souza}
\email{ransell.dsouza@tyndall.ie}
\affiliation{Tyndall National Institute, Lee Maltings, Dyke Parade, Cork T12 R5CP, Ireland}
\author{Jos\'{e} D. Querales-Flores}
\affiliation{Tyndall National Institute, Lee Maltings, Dyke Parade, Cork T12 R5CP, Ireland}
\author{Jiang Cao}
\affiliation{Institut f\"ur Integrierte Systeme, ETH Z\"urich, R\"amistrasse, 101, 8092, Z\"urich, Switzerland}
\alsoaffiliation{School of Electronic and Optical Engineering, Nanjing University of Science and Technology, Nanjing 210094, China}
\author{Stephen Fahy}
\affiliation{Department of Physics, University College Cork, College Road, Cork T12 K8AF, Ireland}
\author{Ivana Savi\'c}
\affiliation{Tyndall National Institute, Lee Maltings, Dyke Parade, Cork T12 R5CP, Ireland}
\email{ivana.savic@tyndall.ie}


\title{Temperature induced band convergence, intervalley scattering and thermoelectric transport in $p$-type PbTe}%\footnote{A footnote for the title}}

\let\oldmaketitle\maketitle
\let\maketitle\relax

\begin{document}

\twocolumn[
\begin{@twocolumnfalse}
\oldmaketitle
\begin{abstract}
Achieving high valley degeneracy (i.e. ``band convergence'') in a material usually results in considerably enhanced thermoelectric properties. However, it is still unclear why this strategy of designing efficient thermoelectric materials is so successful, since the benefit of increased density of states may be severely degraded by intervalley scattering. Using first principles calculations, we investigate these effects in $p$-type PbTe, where temperature induces alignment of the $L$ and $\Sigma$ valleys at $\sim$~620~K. We explicitly show that the thermoelectric power factor and figure of merit peak near the band convergence temperature. The figure of merit maximum is larger than those of the individual $L$ and $\Sigma$ valleys. Surprisingly, intervalley scattering does not considerably affect the figure of merit near the band convergence temperature and optimal doping conditions, although it reduces the power factor by almost a factor of 2.  %\st{near optimal doping conditions.} 
Our results suggest that band convergence will significantly increase the figure of merit if intervalley scattering is roughly proportional to the density of states and the lattice thermal conductivity is considerably lower than the electronic thermal conductivity, even if intervalley scattering is strong.
\end{abstract}
\end{@twocolumnfalse}
]
% Abstract should be written in the present tense and impersonal style (i.e., avoid we), and be at most 200 words long
\section{Introduction}\label{sec:int}
The topological Lifshitz transition is characterized by the evolution or elimination of the Fermi surfaces caused by extrinsic parameters \cite{lifshitz60,volovik17,nishimura19}.
Since thermoelectric (TE) transport coefficients are strongly influenced by the Fermi surface alterations, the Lifshitz transitions have the potential to significantly increase the TE figure of merit $ZT$ \cite{lifshitz60,volovik17,nishimura19}.
Obtaining multiple valley degeneracy in semiconductors (so called ``band convergence") by varying temperature, pressure or alloying is a particular type of the Lifshitz transition. It is also one of the most advantageous methods to enhance $ZT$ that has been used in a number of thermoelectric materials \cite{pei2011,pei12,liu12,zhao13,wang14,tan14,tan15,tan15_hg,tang15,banik15,li16,tan16,wang16,kim17,zheng18,liu18}.
The figure of merit of a material reads
\begin{equation}\label{eq:zt}
ZT = \frac{S^2\sigma}{\kappa_e + \kappa_L} T,
\end{equation}
where $S$, $\sigma$, $\kappa_e$, $\kappa_L$ and $T$ are the Seebeck coefficient, electrical conductivity, electronic thermal conductivity, lattice thermal conductivity and temperature, respectively~\cite{snyder08}. It is believed that an increased density of states due to band convergence increases both $S$ and $\sigma$, thus improving $ZT$ \cite{pei2011,pei12,liu12,zhao13,wang14,tan14,tan15,tan15_hg,tang15,banik15,li16,tan16,wang16,kim17,zheng18,liu18}.

However, the role of intervalley scattering in band convergence strategies to improve $ZT$ is still not well understood~\cite{hong16,Kumarasinghe19,murphy-armando19,park2020,Graziosi2020,park2021,zhou2022}. Band convergence increases the strength of intervalley scattering, thus reducing electrical conductivity. Moreover, according to the Mott relation \cite{xia19}, the energy dependence of intervalley scattering could either enhance or reduce the Seebeck coefficient. Therefore, intervalley scattering is usually considered to be detrimental for the power factor $S^2\sigma$ and $ZT$~\cite{park2020}. Indeed, it has been shown both experimentally and theoretically that band convergence is not beneficial for the TE performance of several materials, such as Zintl CaMg$_{2-x}$Zn$_x$Sb$_2$ alloys~\cite{park2020}.

In this paper, we elucidate the impact of band convergence and intervalley scattering on the thermoelectric properties of $p$-type PbTe using first principles calculations. PbTe and its alloys with PbSe are some of the first materials where the phenomenon of band convergence was exploited to increase $ZT$ \cite{pei2011,pei12}. PbTe has a relatively complicated valence band structure with hole pockets at the L points and along the $\Sigma$ directions in the Brillouin zone~\cite{parker13} (see Fig.~\ref{fig:bs}~(a)). At low temperatures, the highest valence band maxima are at $L$, but as the temperature rises, the valence band maxima at $\Sigma$ become aligned with those at $L$ (at $\sim 620$~K \cite{jose19}), increasing the valley degeneracy \cite{pei2011,pei12,madison2020}. Standard first principle methods to simulate electronic and thermoelectric transport \cite{sohier14,qiu15,li15,fiorentini16,zhou16,gunst16,liu17,ponce18,brunin20,ponce21,zhou21} do not account for such temperature renormalization of the electronic band structure. Recently, we developed a first principles based model of thermoelectric transport in $p$-type PbTe, which explicitly includes those temperature effects~\cite{rd20}. We applied this model to investigate the TE properties of $p$-type PbTe at 300 K~\cite{rd20}, which is far from the band convergence temperature.

Here we use our model from Ref.~\cite{rd20} to understand the influence of the temperature induced band convergence on the thermoelectric transport properties of $p$-type PbTe. Our calculated TE coefficients are in excellent agreement with available experimental data. We demonstrate that both the power factor and TE figure of merit have their maxima near the band convergence temperature. The $ZT$ values near band convergence are larger than those of the individual $L$ and $\Sigma$ pockets at optimal doping concentrations. Scattering between the $L$ and $\Sigma$ valleys, which is approximately proportional to the density of states, significantly decreases both the electrical conductivity and electronic thermal conductivity, and increases the Seebeck coefficient. As a result, intervalley scattering considerably reduces the power factor (by almost a factor of 2) near the band convergence temperature and optimal doping concentrations. Nevertheless, intervalley scattering does not affect the $ZT$ considerably at those temperatures and doping concentrations since the lattice thermal conductivity of PbTe is considerably lower than the electronic thermal conductivity. We thus conclude that an increased density of states due to band convergence is the primary effect driving the $ZT$ increase if intervalley scattering is approximately proportional to the density of states and the lattice thermal conductivity is low. The condition that intervalley scattering is roughly proportional to the density of states is satisfied when the valley minima with the same energy occur at distant points in reciprocal space, as in the case of $p$-type PbTe, or when the valley minima with the same energy occur at the same point in reciprocal space and the dominant scattering mechanism between them is acoustic or non-polar optical phonon scattering.

\section{Methods}
\subsection{Electronic band structure}\label{sec:kane}
The Kane model, derived from the $\bm{k\cdot p}$ Hamiltonian \cite{kane57}, describes the dispersion of the conduction and valence bands for the four $L$ and twelve $\Sigma$ valleys of $p$-type PbTe as follows \cite{rd20}:
\begin{eqnarray}\label{eq:kane_L}
\frac{\hbar^2}{2}\bigg[\frac{(k^L_{\parallel})^2}{m^L_\parallel} + \frac{(k^L_{\perp})^2}{m^L_\perp} \bigg] = E\left(1+\frac{E}{E_g^{L}}\right), %\gamma_{L}(E), 
\end{eqnarray}
\begin{eqnarray}\label{eq:kane_sig}
\frac{\hbar^2}{2}\bigg[\frac{(k^{\Sigma}_{\parallel})^2}{m^{\Sigma}_\parallel} + \frac{(k^{\Sigma}_{\perp_{xy}})^2}{m^{\Sigma}_{\perp_{xy}}} + \frac{(k^{\Sigma}_{\perp_{z}})^2}{m^{\Sigma}_{\perp_{z}}} \bigg]\squeeze =\squeeze E\left(1+\frac{E}{E_g^{\Sigma}}\right),
\end{eqnarray}
where $E$ is the energy of the electronic states with respect to the relevant valley band extrema, $E_g$ is the electronic band gap, and $\hbar$ is the reduced Planck constant.
$m_{\parallel(\perp)}^{L(\Sigma)}$ is the effective mass of the $L(\Sigma)$ valleys along the parallel ($\parallel$)/perpendicular ($\perp$) direction of the wave vector component $k_{\parallel(\perp)}^{L(\Sigma)}$. At $0$~K, we fit the Kane model to the band structure calculated using density functional theory (DFT) and obtain the effective masses of the $L$ and $\Sigma$ valleys. Details of our DFT band structure calculations and all parameters of the electronic band structure at 0 K are given in Ref. \cite{rd20}. The temperature dependence of the band gap and the energy differences between the valence band maxima of the $L$ and $\Sigma$ valleys are taken from our earlier calculations \cite{jose19}. The calculated value of the temperature where the $L$ and $\Sigma$ valleys become aligned is $\sim 620$~K \cite{jose19}. The temperature renormalization of the electronic states was calculated taking into account the thermal expansion and electron-phonon interaction contributions. The latter contribution was computed using the Allen-Heine-Cardona method \cite{allen76,allen81,allen83} and density functional perturbation theory (DFPT)~\cite{gonze97,baroni01}.  The effective masses are taken to be proportional to the electronic band gap \cite{madelung87,yu05, ridley99}, which was justified and verified in our previous work~\cite{rd20}. Therefore, the temperature dependence of the effective masses can be obtained as $m_{\parallel,\perp}^{L}(T)/m_{\parallel,\perp}^{L}(0) = E^{L}_g(T)/E^{L}_g(0)$ \cite{rd20,jiang18,jiang19}. The effective masses of the $\Sigma$ valleys do not change much with temperature due to the large band gap at $\Sigma$, and this effect was neglected in our model~\cite{rd20}.

\begin{figure}[t]
\begin{centering}
\includegraphics[keepaspectratio, width=0.5\textwidth]{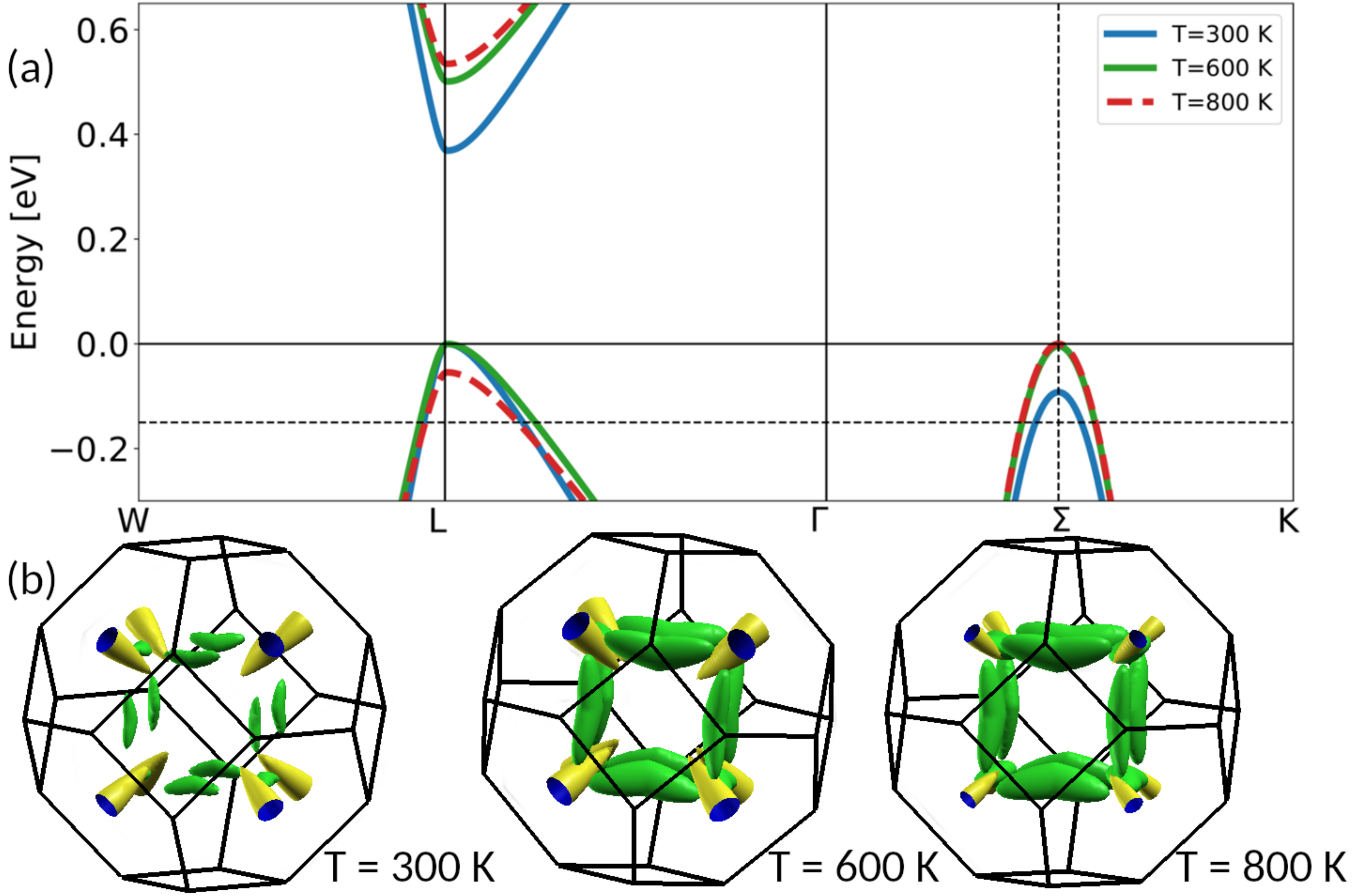}
\caption{\label{fig:bs} {\bf Temperature dependence of the electronic band structure of $p$-type PbTe: }(a) The electronic band structure along selected high symmetry directions in the Brillouin zone at various temperatures. The solid horizontal line at $E = 0$ eV marks the highest valence band maxima (VBM). The $L$ and $\Sigma$ valleys are almost aligned at $600$~K. This effect is known as band convergence. The dotted horizontal line marks the energy value of VBM-0.15 eV, at which the energy iso-surfaces were calculated. (b) Energy iso-surfaces for the $L$ and $\Sigma$ valleys (yellow and green, respectively) at VBM-0.15 eV. }
\end{centering}
\end{figure}

\begin{figure}[t]
\begin{centering}
\includegraphics[keepaspectratio, width=0.45\textwidth]{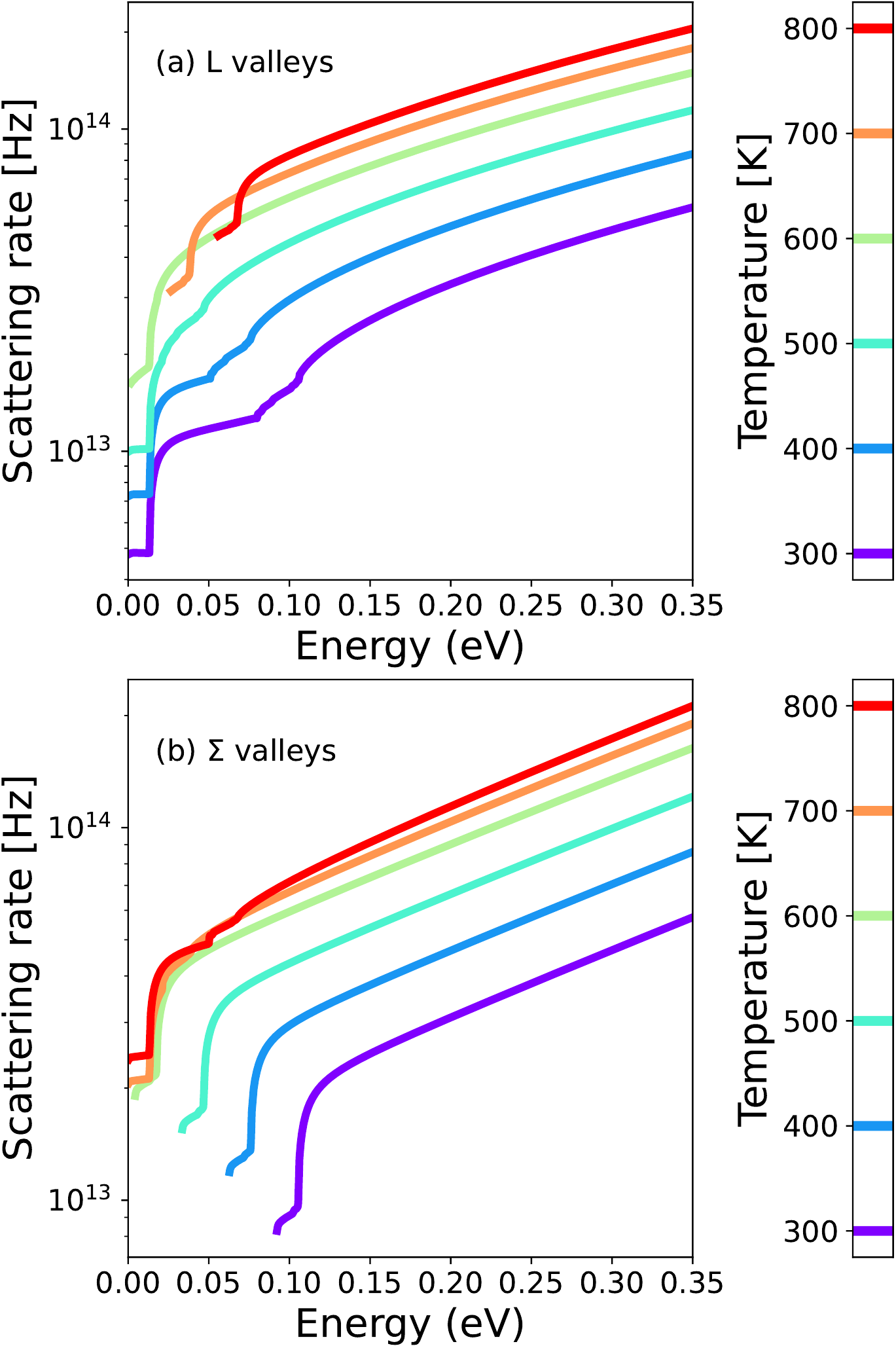}
\caption{\label{fig:SR} {\bf Valley-resolved scattering rates of $p$-type PbTe:} Scattering rates as a function of the absolute value of the hole energy with respect to the highest valence band maxima at various temperatures for (a) the L valleys and (b) the $\Sigma$ valleys.}
\end{centering}
\end{figure}

\subsection{Scattering rates and relaxation times}
The scattering rates and momentum relaxation times due to electron-phonon coupling are calculated from first order perturbation theory. They are functions of the overlap integral between initial and final electronic states, phonon frequencies and coupling coefficients between two electronic states due to a phonon mode \cite{ridley99}.
Standard methods to calculate electron-phonon scattering from first principles usually calculate the scattering rates in the entire Brillouin zone~\cite{ponce18,brunin20,ponce21,zhou21}, but do not account for the temperature dependence of the electronic band structure. However, the scattering events due to long wavelength phonons (intravalley scattering in the case of $p$-type PbTe) or zone edge phonons (intervalley scattering for $p$-type PbTe) are dominant due to their energy and momentum conservation \cite{ridley99,rd20,jiang18}. Our model includes only these contributions.

In our model, the momentum relaxation time of the valence or conduction band state with the wave vector $\bm k$ due to intravalley scattering is expressed as \cite{ridley99}:
\begin{eqnarray}\label{eq:rt}
\tau^{-1}(\bm{k}) \Squeeze &=& \Squeeze \frac{2\pi}{\hbar}\sum_{\bm{q}}\sum_{\lambda} |M_{\lambda,{\bm q}}|^2 \big(1-\cos\theta_{\bm{k'}}\big) \delta_{\bm{k'},\bm{k \pm q}} \nonumber \\ 
&\times& \Squeeze \frac{1-f^0(E_{\bm{k'}})}{1-f^0(E_{\bm{k}})} \delta(E_{\bm k'}-E_{\bm k}\mp \hbar \omega_{\lambda,{\bm q}}),
\end{eqnarray}
where  $f^0$ is the Fermi-Dirac distribution. $|M_{\lambda,{\bm q}}|$ is the electron-phonon matrix element for the phonon mode $\lambda$ with the wave vector $\bm q$ and the phonon frequency $\omega_{\lambda}$. The angle between two Bloch states with the wave vectors $\bm{k}$ and $\bm{k'}$ is denoted by $\theta_{\bm k'}$.
The absorption (emission) of a phonon is represented by the upper (lower) sign.
The Kronecker symbol $\delta_{\bm{k'},\bm{k \pm q}}$ represents the crystal momentum conservation. The scattering rate for the valence  or conduction band state with the wave vector $\bm k$ due to intravalley scattering can be obtained by excluding the term $1-\cos\theta_{\bm{k'}}$ in Eq.~\ref{eq:rt}. 

The momentum relaxation time of the valence band state of the valley $i$ with the wave vector $\bm k$ due to intervalley scattering into the valley $j$ via the phonon mode $\lambda$ is given as \cite{ridley99}:
\begin{eqnarray}
\label{eq:inter}
(\tau^{ij}_\lambda({\bm k}))^{-1}
\Squeeze &=&\Squeeze \frac{\pi V (\Xi_{ij}^{\lambda})^2}{m \omega_{\lambda}} \bigg[n(\omega_{\lambda})D_j(E_{\bm k} - \Delta E_{ij} + \hbar\omega_\lambda)  \nonumber \\
&+& \Squeeze(n(\omega_{\lambda})+1)D_j(E_{\bm k} - \Delta E_{ij} - \hbar\omega_\lambda) \bigg],
\end{eqnarray}
where $V$ and $m$ are the volume and mass of the atoms in the unit cell. $D_j(E)$ is the density of states (DOS) of the valley $j$. $\Delta E_{ij}$ and $\Xi_{ij}$ are the energy difference and deformation potential between the valence band maxima of the valleys $i$ and $j$, respectively. $n(\omega_{\lambda,{\bm q}})$ is the Bose-Einstein distribution. The scattering rate for the valence band state of the valley $i$ with the wave vector $\bm k$ is inversely proportional to the relaxation time given by Eq.~\ref{eq:inter} summed over all the valleys $j$.

Intravalley and intervalley scattering rates have contributions from acoustic, transverse optical and longitudinal (LO) phonon modes. All phonon frequencies and deformation potentials were calculated using DFPT, as explained and tabulated in Ref.~\cite{rd20}. We have not taken into account the temperature dependence of phonon frequencies and deformation potentials, as justified in detail in Ref.~\cite{rd20}. The polar contribution of the LO phonon mode was calculated using the Fr\"ohlich model \cite{jiang18}. %\st{In addition, we accounted for ionized impurity scattering using the Brooks-Herring model} \cite{jiang18}. 
In addition, we explicitly accounted for ionized impurity scattering using the Brooks-Herring model~\cite{jiang18}. We calculated relaxation times due to ionized impurities at each doping concentration and temperature. Ionized impurity scattering has a very small influence on the thermoelectric transport properties of PbTe due to its very large static dielectric constant ($\epsilon_s=356.8$) \cite{rd20, jiang19}.

\subsection{Thermoelectric transport coefficients}
We use the Boltzmann transport theory in the momentum relaxation time approximation to obtain the thermoelectric transport parameters. Within the Kane model, we define the thermoelectric transport kernel function for each valley ($L$ or $\Sigma$) and carrier type (holes or electrons) as \cite{jiang19, rd20}:
\begin{eqnarray}
 L_{\beta}\Squeeze &=& \Squeeze \frac{e^{2-\beta}m_d^{3/2}}{\pi^2}\Squeeze\int_0^\infty \Squeeze\frac{-\partial f^0}{\partial E}\tau \overline{v}^2 (E-E_F)^{\beta} k^{2}dk \label{eq:trans_kernel}, \nonumber \\
m^L_d\Squeeze &=&\Squeeze [m^L_{\parallel}(m^L_{\perp})^2]^{\frac{1}{3}} ; m^{\Sigma}_d = (m^{\Sigma}_{\parallel}m^{\Sigma}_{\perp_{xy}}m^{\Sigma}_{\perp_{z}})^{\frac{1}{3}},
\end{eqnarray}
where $e$ is the electronic charge, $\overline{v}$ is the average group velocity, $m_d$ is the DOS effective masses, and $\tau$ is the carrier relaxation time.
The thermoelectric transport parameters for each valley and carrier type can then be expressed as \cite{rd20}:
\begin{eqnarray}\label{eq:L_i}
\sigma = L_0,\ S = \frac{L_1}{TL_0},\  \kappa_0 &=& \frac{L_2}{T}.
\end{eqnarray}
The total values of the thermoelectric transport parameters in our three band model, including the valence and conduction bands at the $L$ valleys and the valence band at the $\Sigma$ valleys, can then be written as \cite{rd20}:

\begin{eqnarray}\label{eq:sigma}
\sigma_{tot} \Squeeze &=& \Squeeze N_L\sigma_{h_L} + N_{\Sigma}\sigma_{h_{\Sigma}} + N_L\sigma_{e_L}, \\
  S_{tot} \Squeeze &=& \Squeeze \frac{N_L S_{L_e} \sigma_{L_e} \squeeze+\squeeze N_L S_{L_h} \sigma_{L_h} \squeeze+\squeeze N_{\Sigma} S_{{\Sigma}_h} \sigma_{{\Sigma}_h}}{\sigma_{tot}}, \\
  %\kappa_{e_{tot}} \Squeeze &=&\Squeeze N_L\kappa_{0,h_L}\squeeze + \squeeze N_{\Sigma}\kappa_{0,h_{\Sigma}}\Squeeze + \squeeze N_L\kappa_{0,e_L} \squeeze - \squeeze S_{tot}^2\sigma_{tot}T.
  \kappa_{e_{tot}} &=& N_L\kappa_{0,h_L} + N_{\Sigma}\kappa_{0,h_{\Sigma}} + N_L\kappa_{0,e_L}  \nonumber \\&-&  S_{tot}^2\sigma_{tot}T.
\end{eqnarray}
where $N_L = 4$ and $N_{\Sigma} = 12$ are the number of the $L$ and $\Sigma$ valleys, respectively. The subscripts $e$ and $h$ represent the contributions from electrons and holes, respectively.
The lattice thermal conductivity used in the present computation of $ZT$ was taken from our earlier first principles calculations~\cite{murphy16}.

\section{Results and discussion}
\subsection{Temperature dependence of electronic bands and transport properties}
\begin{figure*}[t]
\begin{centering}
\includegraphics[keepaspectratio, width=1\textwidth]{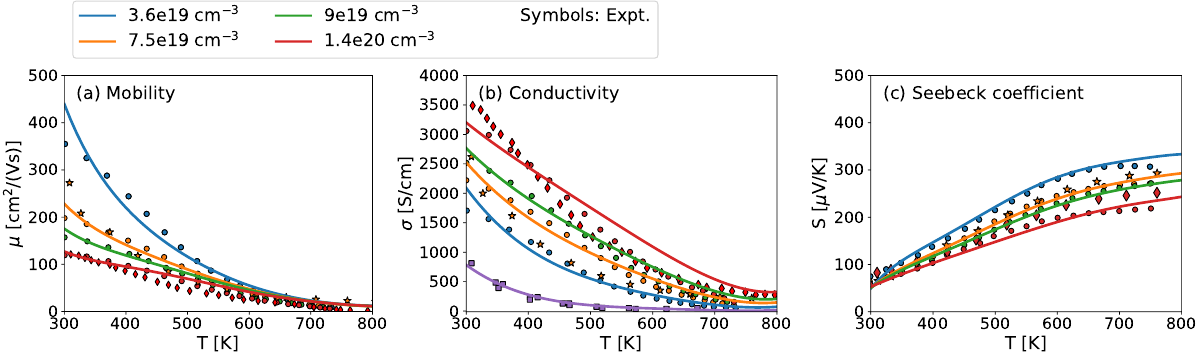}
\caption{\label{fig:TP} {\bf Thermoelectric transport coefficients of $p$-type PbTe:} (a) Mobility, (b) conductivity and (c) the Seebeck coefficient versus temperature at various carrier doping concentrations. %\st{and temperatures.} 
The solid lines represent our calculations. The circles and squares refer to the experimental data for Na-doped PbTe, taken from Refs.~\cite{pei11, vineis08}.  The star and diamond symbols refer to the experimental data for 1\% ($\sim 6.3\times 10^{19}$ cm$^{-3}$) and 2\% ($\sim 1.4\times 10^{20}$ cm$^{-3}$) Na-doped PbTe, taken from Refs.~\cite{girard11, jood20}.} 
\end{centering}
\end{figure*}
First we illustrate the temperature dependence of the electronic band states relevant for thermoelectric transport in $p$-type PbTe. In Fig.~\ref{fig:bs}~(a) we plot the computed electronic band structure of PbTe along the high symmetry directions in the first Brillouin zone  at $300$ K (blue), $600$~K (green) and $800$ K (red).
Fig.~\ref{fig:bs}~(b) shows the calculated energy iso-surfaces at those temperatures for the energy $0.15$ eV below the highest valence band maxima (VBM).
The pockets shown in yellow correspond to the eight half pockets at the $L$ points. The pockets shown in green correspond to the twelve $\Sigma$ valleys, which have much larger effective masses than the $L$ valleys~\cite{rd20}. 
At $300$~K, the energy difference between the $L$ and $\Sigma$ valleys is relatively large ($\Delta E_{L-\Sigma} = E_{\Sigma}-E_L = 0.092$~eV), indicating that electronic conduction occurs mainly through the $L$ valleys. At $620$ K, $\Delta E_{L-\Sigma} = 0$~eV and the $L$ and $\Sigma$ valleys are aligned, i.e. band convergence takes place. At $800$\,K, $\Delta E_{L-\Sigma} = -0.054$~eV, indicating that the highest VBM are located at the $\Sigma$ valleys, which mostly determine the transport coefficients at high temperatures.

Next we discuss the non-linear temperature variations of the electronic scattering rates induced by the temperature dependence of the electronic band structure. Figs. \ref{fig:SR} (a) and (b) show the calculated total scattering rates for $p$-type PbTe of the $L$ and $\Sigma$ valleys, respectively, for various temperatures ranging from $300$~K to $800$~K. We plot the scattering rates versus the absolute value of the hole energy with respect to the highest VBM, labeled as $E$. As a result, the total scattering rates of the $L$ valleys appear at $E = 0$ for $T<620$~K where the highest VBM is at L, while they appear at $E > 0$ for $T>620$~K since the highest VBM is at the $\Sigma$ valleys. The opposite trend is seen for the $\Sigma$ valleys. Near $E=\Delta E_{L-\Sigma}$, there is a sudden increase in the scattering rates because of intervalley scattering. We also find a ripple like energy dependent behavior near $E=\Delta E_{L-\Sigma}$ due to emission of different phonon modes. All these features of the scattering rates are strongly dependent on temperature.

Our analysis shows that the intravalley scattering rates of the $L$ and $\Sigma$ valleys are larger than the corresponding intervalley scattering rates. Both intravalley and intervalley scattering in both types of valleys are dominated by longitudinal optical phonons.  The intervalley scattering rate from the $L$ valleys into the highly degenerate and heavy $\Sigma$ valleys is larger than the intervalley scattering rate from the $\Sigma$ valleys into the other $\Sigma$ and $L$ valleys. We have shown previously that the intervalley scattering between the L points is forbidden by symmetry \cite{rd20}. Therefore, the scattering between the L valleys is considered to be much smaller than that between the $L$ and $\Sigma$ valleys, and is neglected in our model \cite{rd20}.

Next we calculate the mobility, conductivity and Seebeck coefficient,
%using the momentum relaxation times derived from these scattering rates,
%(see Methods),
and compare our results with available experimental results for a wide range of temperatures and doping concentrations \cite{pei11}, see Fig.~\ref{fig:TP}. 
We also present our calculated resistivity values in the \hyperref[SI]{Supporting Information} (see Fig.~S1).
Experimentally, divalent Pb atoms can be substituted with monovalent Na atoms to obtain $p$-type PbTe \cite{pei11}. Na-doped PbTe does not have resonant states in the valence band structure \cite{chernik68} and the related measurements should agree well with our calculations. Indeed, it is evident from Fig. \ref{fig:TP} that our results are in very good agreement with the experimental results on Na-doped PbTe for various doping concentrations~\cite{pei11}.

Our results for conductivity and mobility at lower temperatures and doping concentrations are in better agreement with experiments than those obtained at higher temperatures. This may be due to the fact that the Kane model
%we used (see Methods)
captures the electronic band structure very accurately only at lower energies~\cite{rd20}. Furthermore, while the measurements on Na-doped PbTe in Ref. \cite{pei11} do not show error bars, Vineis {\it et al.} \cite{vineis08} have shown that the error bars for the measured conductivity, Seebeck coefficient, and power factor are rather large for $p$-type PbTe.%\st{Moreover, $p$-type PbTe can be realized by substituting Pb atoms with various dopant atoms %such as Na and Mg} \cite{pei11_sc, pei11,girard11,jood20}, \st{which have slightly different %values of the same thermoelectric parameters.} 
Therefore, taking into consideration the size of the experimental error bars %\st{, the effects of different dopants} 
and the approximations in our model, we deem our calculations to be in very good agreement with experiments.

\subsection{Valley and intervalley scattering contributions to thermoelectric parameters}

\begin{figure}[t]
%\begin{centering}
\includegraphics[keepaspectratio, width=0.5\textwidth, center]{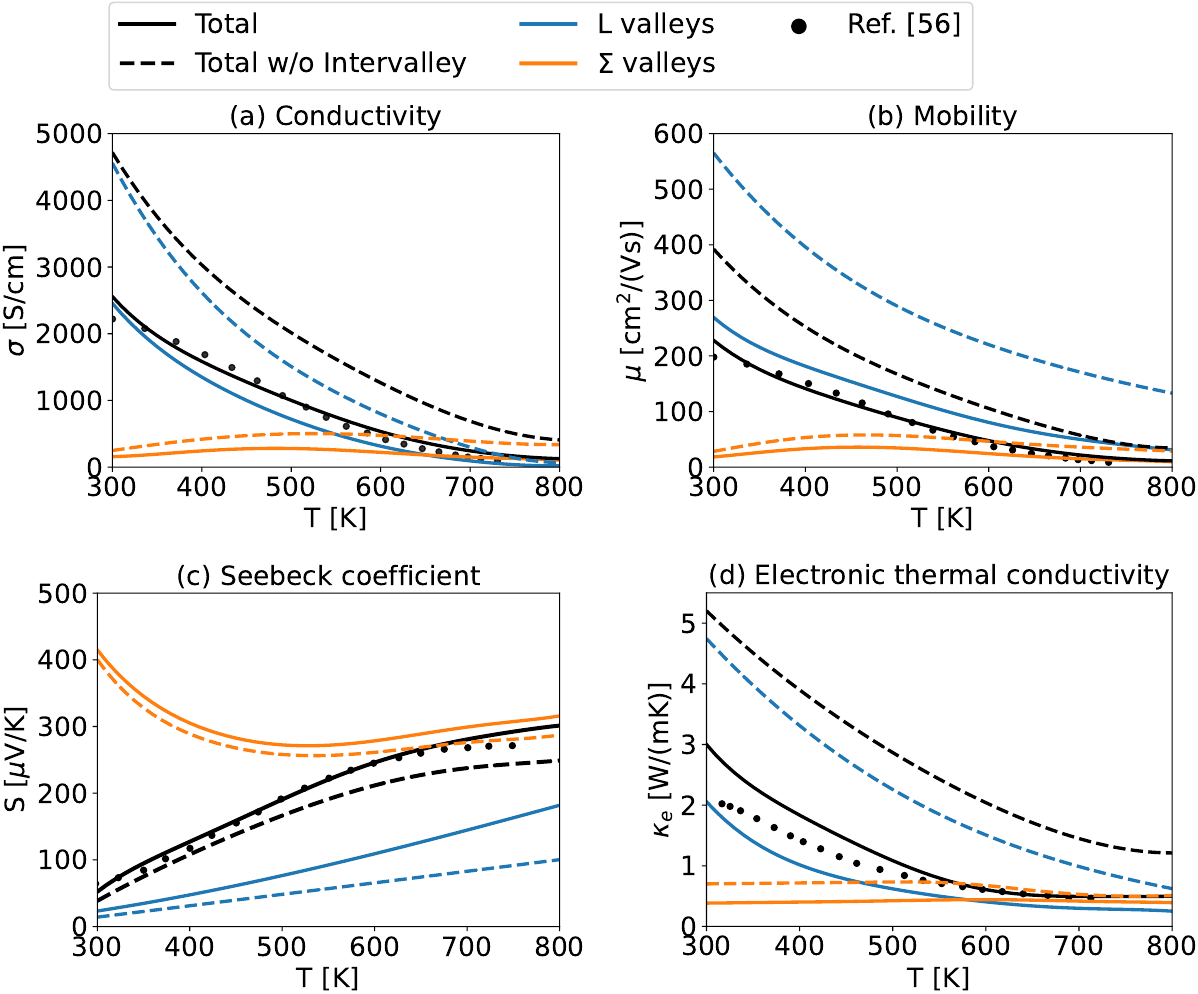}
\caption{\label{fig:TP_const_n} {\bf Valley-resolved thermoelectric transport coefficents of $p$-type PbTe:} (a) Conductivity, (b) mobility, (c) the Seebeck coefficient and (d) electronic thermal conductivity versus temperature for the doping concentration of $7.5\times10^{19}$ cm$^{-3}$. The solid (dashed) lines represent our calculations including (excluding) intervalley scattering. The blue curves correspond to the $L$ valleys (both valence and conduction bands) and the orange curves correspond to the $\Sigma$ valleys. The circles refer to the experimental data on Na-doped PbTe with the doping concentration of $7.5\times10^{19}$ cm$^{-3}$, taken from Ref. \cite{pei11}.
}
%\end{centering}
\end{figure}

Here we investigate the effect of band convergence and intervalley scattering on each of the thermoelectric parameters. In Fig.~\ref{fig:TP_const_n}, we plot (a) the conductivity, (b) the mobility,  (c) the Seebeck coefficient, and (d) the electronic thermal conductivity of $p$-type PbTe as a function of temperature for the doping concentration of $7.5 \times 10^{19}$ cm$^{-3}$. Our results including (excluding) intervalley scattering are plotted as the solid (dashed) black line. The solid blue (orange) lines correspond to the results including only the $L$ $(\Sigma)$ valleys with intervalley scattering and the dashed lines are for the same without intervalley scattering. The contribution of the $L$ valleys to thermoelectric transport properties contains the contributions from both conduction and valence bands. However, the contribution of the $L$ valleys electrons to thermoelectric transport is very small even at high temperatures (see Fig.~S2 in the \hyperref[SI]{Supporting Information}). This is due to the sizeable band gap (larger than $0.45$ eV for temperatures above 600~K), the vicinity of the Fermi level to the valence band maxima (see Fig.~S3 in the \hyperref[SI]{Supporting Information}), and the small density of the conduction band states compared to that of the valence band states. For all thermoelectric transport parameters, it is evident from Fig.~\ref{fig:TP_const_n} (a) that as the temperature is raised from $300$\,K to $800$\,K, the primary transport channels for electronic conduction change from the $L$ valleys to the $\Sigma$ valleys.

As expected, intervalley scattering causes a reduction in the total conductivity and mobility, as well as their contributions from the $L$ and $\Sigma$ valleys, see Figs. \ref{fig:TP_const_n} (a) and (b). The decrease in conductivity with temperature due to intervalley scattering is more pronounced for the $L$ valleys since the intervalley scattering from the L valleys into the highly degenerate and heavy $\Sigma$ bands is very strong.
%Therefore, the conductivity of the lighter $L$ valleys decreases much more than that of the heavier $\Sigma$ valleys because of the comparatively larger intervalley scattering.

In contrast, intervalley scattering increases the Seebeck coefficient for both types of valleys, as well as the total Seebeck coefficient, see Fig.~\ref{fig:TP_const_n}~(c). 
%\st{The opposite effects of intervalley scattering on the Seebeck coefficient and the conductivity can be expected since the Seebeck coefficient is inversely proportional to the conductivity, see Eq.}~\ref{eq:L_i}. 
%\st{Furthermore,  the Seebeck coefficient enhancement due to intervalley scattering can be understood from the Mott expression for the Seebeck coefficient, $S \sim \left.\left (\frac{1}{D}\frac{\text{d}D}{\text{d}E}+\frac{1}{v}\frac{\text{d}v}{\text{d}E}+\frac{1}{\tau}\frac{\text{d}\tau}{\text{d}E}\right)\right|_{E=E_F}$, where $D$, $v$, $\tau$ and $E_F$ are the density of states (DOS), group velocities, momentum relaxation times and the Fermi level, respectively}~\cite{xia19}. 
%\st{Convergence of the $L$ and $\Sigma$ valleys causes a sudden increase in the DOS, leading to the large positive term $\frac{1}{D}\frac{\text{d}D}{\text{d}E}$. Intervalley scattering between distant valleys in reciprocal space is approximately proportional to the DOS (see Eq.}~\ref{eq:inter}), 
%\st{which also results in the large positive term $\frac{1}{\tau}\frac{\text{d}\tau}{\text{d}E}$.} 
The Seebeck coefficient enhancement due to intervalley scattering can be understood from the energy dependence of the integrands in Eq. \ref{eq:trans_kernel}. The integrand of $L_0$ is proportional to the derivative of the Fermi-Dirac distribution with respect to energy ($\frac{-\partial f^0}{\partial E}$), while the integrand of $L_1$ is proportional to $(E-E_F)\frac{-\partial f^0}{\partial E}$. Both integrands are proportional to the DOS, group velocities and relaxation times, and therefore equal to zero at the band edge. The integrand of $L_1$ approaches the zero value at the band edge at a much faster rate in comparison to $L_0$ since $\frac{-\partial f^0}{\partial E}$ has its maximum at the Fermi level (which is near the band edge for higher temperatures and doping concentrations, see Fig. S2 in the \hyperref[SI]{Supporting Information}), while $(E-E_F)\frac{-\partial f^0}{\partial E}$ is equal to zero at the Fermi level. As a result, when intervalley scattering is approximately proportional to the DOS, we find that the decrease of $L_0$ due to intervalley scattering will be larger than the decrease of $L_1$, thus enhancing the Seebeck coefficient.
Therefore, an increase in the energy dependence of the DOS due to band convergence directly increases the energy dependence of intervalley scattering rates and momentum relaxation times, and both effects enhance the Seebeck coefficient. This and the following conclusions might not apply to materials where intervalley scattering is not proportional to the density of states, which can happen e.g.~when the degenerate valleys relevant for transport are located at the same point in reciprocal space and  the scattering between them is predominantly by polar optical phonons~\cite{park2021}. 

We note that the electrical conductivity and the Seebeck coefficient of the $\Sigma$ valleys have an unusual temperature dependence, as shown in Figs.~\ref{fig:TP_const_n}~(a) and (c). At lower temperatures, the $\Sigma$ valleys are positioned further away from the Fermi level compared to the $L$ valleys. This causes the conductivity (the Seebeck coefficient) of the $\Sigma$ valleys to be much smaller (larger) than that of the $L$ valleys. As temperature increases, the conductivity of the $\Sigma$ valleys initially increases due to the increased carrier concentration of the $\Sigma$ valleys, while the Seebeck coefficient decreases. Near the band convergence temperature, the $\Sigma$ valleys begin to dominate electronic conduction, and the temperature dependence of the conductivity and Seebeck coefficient of the $\Sigma$ valleys becomes similar to those of the $L$ valleys.

Fig. \ref{fig:TP_const_n} (d) illustrates that the increase in scattering rates due to intervalley scattering causes a large reduction in the electronic thermal conductivity. This is beneficial for increasing $ZT$ since the lattice thermal conductivity of PbTe is lower than than the electronic thermal conductivity for the considered and larger doping concentrations (see Fig. S4 (e) in the \hyperref[SI]{Supporting Information}). For example, our calculated lattice thermal conductivity at 300~K is $2.4$ W/mK~\cite{murphy16}, in good agreement with experiments. Pei {\it et al.} \cite{pei11} estimated the electronic thermal conductivity  of Na-doped PbTe as a function of temperature for various doping concentrations from the total thermal conductivity measurements. For all measured doping concentrations, $\kappa_e$ decreases monotonically for temperatures lower than $600$\,K, while for larger temperatures, $\kappa_e$ has a weak temperature dependence, see Fig. \ref{fig:TP_const_n} (d). Our results can explain this trend. At temperatures larger than the band convergence temperature, the behavior of the electronic thermal conductivity is governed by the $\Sigma$ valleys where $\kappa_e$ does not vary much with temperature, similarly to the electronic conductivity. At lower temperatures, $\kappa_e$ is determined by the $L$ valleys, monotonically decreasing with temperature, in the same manner as the electrical conductivity.

We also show the Lorenz number values, defined as the ratio of electronic thermal conductivity and electrical conductivity, as a function of temperature for the doping concentration of of $7.5 \times 10^{19}$ cm$^{-3}$ in the \hyperref[SI]{Supporting Information} (see Fig.~S4 (f)). The Lorenz number values range between $2 \times 10^{-8}$ V$^2$/K$^2$ and $4 \times 10^{-8}$ V$^2$/K$^2$, and do not deviate substantially from the value given by the Wiedmann-Franz law ($2.44 \times 10^{-8}$ V$^2$/K$^2$). Interestingly, the Lorenz number typically decreases when intervalley scattering is accounted for in the calculation, and reaches a minimum near the band convergence temperature of 620 K. Therefore, the difference in the electronic thermal conductivity values with and without intervalley scattering in Fig. 4 (d) is not only due to the increase of the scattering rates due to intervalley scattering, but also due to their energy dependence.

\begin{figure}[h!]
\begin{centering}
\includegraphics[keepaspectratio, width=0.33\textwidth, center]{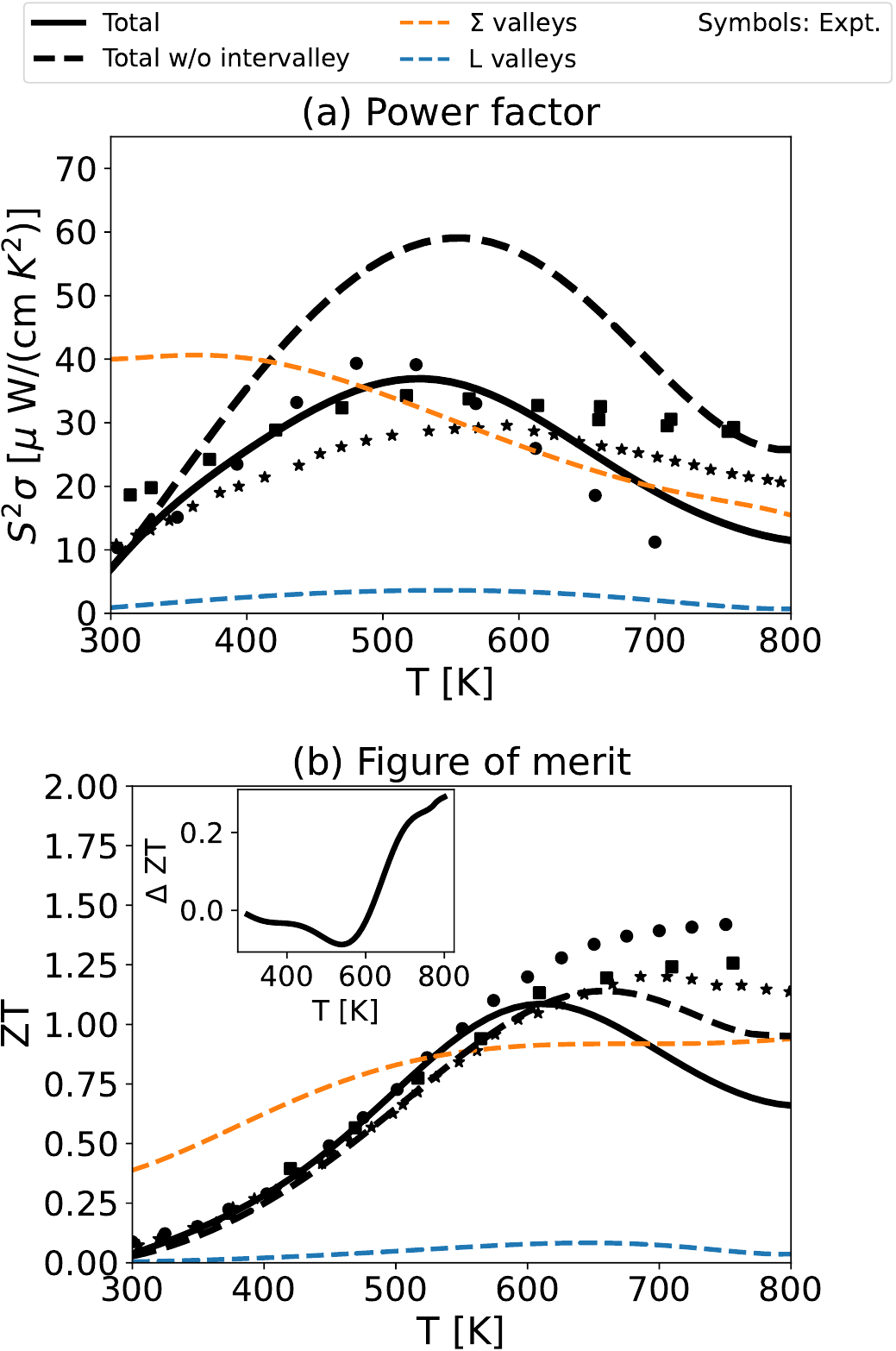}
\caption{\label{fig:PZ_ZT_const_n} {\bf Valley-resolved thermoelectric power factor and figure of merit of $p$-type PbTe:} (a) Power factor and (b) thermoelectric figure of merit  versus temperature  for the doping concentration of $7.5\times10^{19}$ cm$^{-3}$. The solid (dashed) lines represent our calculations including (excluding) intervalley scattering.  The blue curves correspond to the $L$ valleys (both valence and conduction bands) and the orange curves correspond to the $\Sigma$ valleys. The circles refer to the experimental data on Na-doped PbTe for the doping concentration of $7.5\times10^{19}$ cm$^{-3}$, taken from Ref. \cite{pei11}. The star symbols and the squares refer to the experimental data on 1\% and 2\% Na-doped PbTe (corresponding to $\sim 6.3\times 10^{19}$ cm$^{-3}$ and $\sim 1.4\times 10^{20}$ cm$^{-3}$, respectively), taken from Refs. \cite{girard11,jood20}. Inset: Difference between the thermoelectric figure of merit without and with intervalley scattering ($\Delta ZT$) at the doping concentration of $7.5\times10^{19}$ cm$^{-3}$.
}
\end{centering}
\end{figure}

Fig.~\ref{fig:PZ_ZT_const_n} shows the temperature dependence of (a) the power factor and (b) the thermoelectric figure of merit of $p$-type PbTe for the doping concentration of $7.5\times 10^{19}$ cm$^{-3}$, as well as the individual valley contributions. The total power factor and $ZT$ both peak near the band convergence temperature of $620$~K. The total maximal values of these quantities are larger than those of the individual valleys near the band convergence temperature. This is partly because both types of valleys have comparable values of the conductivity at those temperatures, as shown in Fig. \ref{fig:TP_const_n} (a), thus increasing the total conductivity. Furthermore, since the total Seebeck coefficient is governed by the heavy $\Sigma$ valleys, it is not much smaller than the Seebeck coefficient of the $\Sigma$ valleys (see Fig. \ref{fig:TP_const_n} (c)).
%Therefore, the total power factor is larger than each of the individual valleys near the band convergence temperature for the doping concentration of $7.5\times 10^{19}$cm$^{-3}$.

We find that intervalley scattering is detrimental to the power factor near the band convergence temperature, as illustrated in Fig.~\ref{fig:PZ_ZT_const_n} (a). This is because the decrease in conductivity as a result of intervalley scattering (Fig.~\ref{fig:TP_const_n} (a)) is too severe to compensate for the increase in the Seebeck coefficient (Fig.~\ref{fig:TP_const_n} (c)). However, intervalley scattering also considerably reduces the electronic thermal conductivity, not changing the $ZT$ values much. Therefore, intervalley scattering, while having a large influence on the power factor, has only a small effect on the thermoelectric figure of merit when intervalley scattering is approximately proportional to the density of states, and the lattice thermal conductivity is considerably lower than the electronic thermal conductivity.

We have also calculated the change in $ZT$ without and with intervalley scattering for a range of doping concentrations and temperatures, see the inset of Fig.~5 (b) and Figs.~S4 (a)-(c) in the \hyperref[SI]{Supporting Information}. We find that the $ZT$ values are not affected much by intervalley scattering at either lower temperatures (around the band convergence temperature and below) or lower doping concentrations (below $\sim 10^{19}$ cm$^{-3}$). On the other hand, $ZT$ decreases more substantially due to intervalley scattering at higher temperatures and higher doping concentrations. This trend can be understood from the relative increase of the electronic thermal conductivity compared to the lattice thermal conductivity, see Fig. S4 (e) in the \hyperref[SI]{Supporting Information}. At temperatures roughly below the band convergennce temperature, $\kappa_e$ without intervalley scattering is considerably larger than $\kappa_L$. Therefore, the total thermal conductivity is dominated by the electronic contribution in the absence of intervalley scattering. As a result, the effect of the decrease in the power factor due to intervalley scattering on the figure of merit is almost compensated by the decrease in the electronic thermal conductivity caused by intervalley scattering. However, at higher temperatures, $\kappa_e$ without intervalley scattering is more comparable to $\kappa_L$, Fig. S4 (e). When intervalley scattering is taken into account, the electronic thermal conductivity is lower than that of the lattice thermal conductivity. Therefore, the total thermal conductivity is dominated by the lattice contribution. The lowering of the electronic thermal conductivity in the high temperature range does not contribute further to enhancing $ZT$ and hence intervalley scattering is detrimental to the thermoelectric figure of merit at high temperatures.

\subsection{Band convergence and overall thermoelectric performance}
\begin{figure*}[t]
\begin{centering}
\includegraphics[keepaspectratio, width=\textwidth]{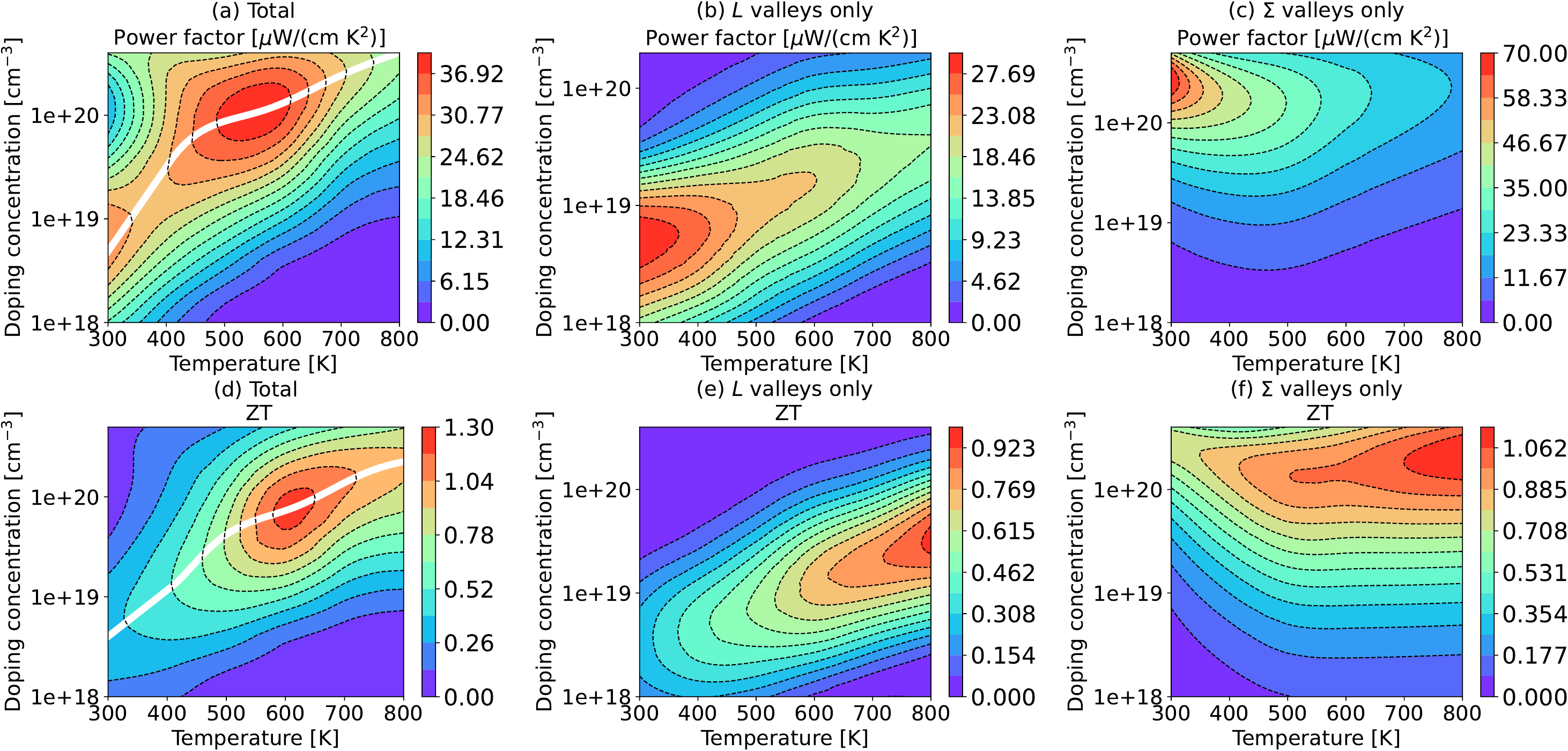}
\caption{\label{fig:PF_ZT_cont} {\bf Thermoelectric power factor and figure of merit of $p$-type PbTe: } Contour plots of the power factor (top panel: (a), (b) and (c)) and thermoelectric figure of merit (bottom panel: (d), (e) and (f)) as a function of temperature and doping concentration. The first column ((a) and (d)) corresponds to the total calculated values when both $L$ and $\Sigma$ valleys are taken into account. The middle column ((b) and (e)) represents the results including the $L$ valleys only. The last column ((c) and (f)) refers to the results including the $\Sigma$ valleys only. The solid white lines represent the optimal doping concentration for maximizing the power factor and figure of merit at a given temperature.}
\end{centering}
\end{figure*}
To examine whether band convergence is beneficial for enhancing the maximal values of the power factor and figure of merit, we calculate these quantities for different doping concentrations and temperatures, see Figs. \ref{fig:PF_ZT_cont} (a) and (d). The white lines show the optimal doping concentration for maximizing the power factor and $ZT$ at each temperature.
%, see the white lines in Figs. \ref{fig:PF_ZT_cont} (a) and (d).
We see that the global maxima of both quantities are obtained near the band convergence temperature, which suggests that band convergence indeed boosts the thermoelectric performance of $p$-type PbTe. Furthermore, the occurrence of the power factor and $ZT$ maxima near the band convergence temperature indicates that these maxima can give a reasonable estimate of the band convergence temperature in $p$-type PbTe.

Figs. \ref{fig:PF_ZT_cont} (b) and (c) show the power factor of the $L$ and $\Sigma$ valleys, respectively, while Figs. \ref{fig:PF_ZT_cont} (e) and (f) show their thermoelectric figure of merit. The total $ZT$ maximum is considerably larger than the $ZT$ maxima of the $L$ and $\Sigma$ valleys, while the maximum of the total power factor is lower than that of the $\Sigma$ valleys. The global maximum of the power factor and $ZT$ of the $L$ valleys is at low and high temperatures, respectively. This behavior is similar to that of $n$-type PbTe as seen experimentally \cite{pei14} and theoretically \cite{jiang19}, since the electronic conduction in $n$-type PbTe is mainly due to the $L$ valleys. The power factor of the $\Sigma$ valleys is very large for extremely high doping and low temperatures, and substantially larger than the maximal value of the total power factor. This is due to the fact that the $\Sigma$ bands are located much further away from the Fermi level for those conditions, thus increasing the Seebeck coefficient substantially. Even though the power factor is larger in the absence of the $L$ valleys, the largest values of $ZT$ are obtained in the presence of both types of valleys near the band convergence temperature, which clearly demonstrates that band convergence is favorable for increasing the figure of merit of $p$-type PbTe. We expect that band convergence will also enhance the figure of merit of other thermoelectric materials with low lattice thermal conductivity and intervalley scattering approximately proportional to the density of states, which occurs when the converging valley minima are either (i) at distant points in recipriocal space, or (ii) at the same point in reciprocal space, and acoustic or non-polar optical scattering between them is the dominant scattering mechanism.

Apart from increasing the figure of merit, the concept of band convergence can also be exploited for reducing the optimal operating temperature of thermoelectric materials. To illustrate this, we note that $n$-type PbTe has a large figure of merit only at very high temperatures ($> 800$ K) \cite{jiang19, pei14}, similarly to the $ZT$ values of the $L$ valleys shown in Fig. \ref{fig:PF_ZT_cont} (e). In contrast, band convergence in $p$-type PbTe leads to the maximal values of $ZT$ for temperatures between $600$\,K and $700$\,K (near the band convergence temperature). Since most of the waste heat generated due to fossil fuels have low grade temperatures ($< 600$\,K) \cite{schierning18}, our calculations clearly show that band convergence is a possible route to optimize the figure of merit for those lower temperatures. Furthermore, achieving band convergence or high valley degeneracy near room temperature in materials other than PbTe would allow the $ZT$ enhancement in this temperature range where it is particularly challenging to increase $ZT$. Indeed, the most efficient thermoelectric materials near 300 K are the alloys of Bi$_2$Te$_3$, which exhibit high valley degeneracy~\cite{witting19}.

\section{Conclusion}
We have investigated the effects of temperature driven band convergence between the $L$ and $\Sigma$ valleys on the thermoelectric transport properties of $p$-type PbTe  from first principles. Our calculated thermoelectric transport coefficients are in excellent agreement with experimental results for a wide range of temperatures and doping concentrations. We find that band convergence is indeed beneficial for increasing the thermoelectric figure of merit, since its maximum appears near the band convergence temperature of $\sim 620$~K and is larger than the corresponding maxima of the individual $L$ and $\Sigma$ valleys. This is mostly a result of an increased density of states due to band convergence. Intervalley scattering does not affect the figure of merit much, since it is proportional to the density of states and the lattice thermal conductivity is appreciably lower than the electronic thermal conductivity. On the other hand, intervalley scattering significantly reduces the power factor near the band convergence temperature and for optimal doping conditions. In addition to increasing the figure of merit, band convergence also helps reducing the optimal operating temperature of $p$-type PbTe.
%down to the band convergence temperature, compared to $n$-type PbTe where the $L$ valleys mostly contribute to thermoelectric transport.

\section*{Acknowledgement}

This project has received funding from the European Union’s Horizon 2020 research and innovation programme under the Marie Skłodowska-Curie grant agreement number 713567. This work is partly supported by Science Foundation Ireland under grant numbers 15/IA/3160 and 13/RC/2077. The later grant is co-funded under the European Regional Development Fund. We acknowledge the Irish Centre for High-End Computing (ICHEC) for the provision of computational facilities.

\section*{Supporting Information Available}\label{SI}
The Supporting Information is available free of charge at [url]. It contains figures showing the resistivity of $p$-type PbTe, the Fermi level positions, the bipolar contribution to thermoelectric transport properties, and the influence of intervalley scattering on the figure of merit, the relative difference between the electrical and lattice thermal conductivity and the Lorenz number.

\bibliographystyle{achemso-demo} 
\bibliography{bc_rd22_acs}

\end{document}